\documentclass{elsarticle}

\journal{HAL} 

\usepackage{hyperref}
\usepackage{amsmath,amsfonts,amssymb,amsthm}
\usepackage{mathtools}
\usepackage{multirow}
\usepackage{graphicx,color}

\newcommand{\G}{\mathbb{G}}
\newcommand{\Z}{\mathbb{Z}}

\newcommand{\F}{\mathbb{F}}

\newcommand{\Ec}{\mathcal{E}}
\newcommand{\W}{\mathcal{W}}
\newcommand{\Ac}{\mathcal{A}}
\newcommand{\Cc}{\mathcal{C}}
\newcommand{\Hc}{\mathcal{H}}
\newcommand{\Jac}{\mathrm{Jac}}

\newcommand{\Pt}{\mathsf{P}}
\newcommand{\Qt}{\mathsf{Q}}
\newcommand{\Dv}{\mathsf{D}}

\newcommand{\Div}{\mathrm{Div}}
\newcommand{\dve}{\mathsf{div}}
\newcommand{\Prin}{\mathrm{Prin}}

\newcommand{\Grp}{\mathcal{G}}

\usepackage{tikz}
\usepackage{tikz-cd}
\usetikzlibrary{matrix}
\usepackage{caption}

\newtheorem{remark}{Remark}
\begin{document}

\begin{frontmatter}

\title{Extending the GLS endomorphism \\ to speed up GHS Weil descent using Magma}

\author[CINVESTAV,TII]{Jes\'us-Javier Chi-Dom\'inguez\corref{mycorrespondingauthor}}
\cortext[mycorrespondingauthor]{Corresponding author}
\address[TII]{Technology Innovation Institute (TII), Abu Dhabi, United Arab Emirates}
\ead{jesus.dominguez@tii.ae}

\author[CINVESTAV,TII]{Francisco Rodr\'iguez-Henr\'iquez\fnref{emailRH}}
\address[CINVESTAV]{Computer Science Department, Center for Research and Advanced Studies of the National Polytechnic Institute of Mexico (Cinvestav - IPN), Mexico City, Mexico}
\fntext[emailRH]{{\it Email address:} {\tt francisco@cs.cinvestav.mx}}

\author[LIX]{Benjamin Smith\fnref{emailS}}
\address[LIX]{Inria and Laboratoire d'Informatique de l'\'Ecole polytechnique (LIX), Institut Polytechnique de Paris, Palaiseau, France}
\fntext[emailS]{{\it Email address:} {\tt smith@lix.polytechnique.fr}}

\begin{abstract}
    Let \(q~=~2^n\), and let \(\Ec / \F_{q^{\ell}}\) be a generalized
    Galbraith--Lin--Scott (GLS) binary curve, with $\ell \ge 2$ and \((\ell, n) = 1\). 
    We show that the GLS endomorphism on \(\Ec / \F_{q^{\ell}}\) 
    induces an efficient endomorphism 
    on the Jacobian \(\Jac_\Hc(\F_q)\) of the genus-\(g\) hyperelliptic curve \(\Hc\) 
    corresponding to the image of the GHS Weil-descent attack applied to
    \(\Ec / \F_{q^\ell}\),
    and that this
    endomorphism yields a factor-$n$ speedup when using standard index-calculus procedures 
    for solving the Discrete Logarithm Problem (DLP) on \(\Jac_\Hc(\F_q)\). Our analysis is backed up by the explicit computation of a discrete logarithm defined on a 
    prime-order subgroup of a GLS elliptic curve over the field
    $\F_{2^{5\cdot 31}}$. A Magma implementation of our algorithm
    finds the aforementioned discrete logarithm in about $1,035$ CPU-days.
\end{abstract}


\end{frontmatter}

\section{Introduction}
Let $\G$ be an additively-written cyclic group of order
$N$. Given an element $\Pt\in\G$ of order $r\mid N$ 
and $\Qt\in \langle \Pt \rangle$, the Discrete Logarithm Problem (DLP)
in~$\G$ is to compute an integer $x$ (if it exists) such that $[x]\Pt = \Qt$. The integer $0
\le x < r$ is called the discrete logarithm 
of $\Qt$ with respect to the base $\Pt$.

In this work, we are interested in the 
case where $\G = \Ec(\F_{q^\ell})$ for an elliptic curve $\Ec$ over a
binary extension field $\F_{q^\ell}$
with $q = 2^n$ and $\ell \ge 2$.
We will be equally interested in the case when $\G$ is the 
Jacobian \(\Jac_\Hc(\F_q)\) of a genus-\(g\) curve $\Hc$ over $\F_q$, and $\Pt, \Qt$ 
are divisors belonging to \(\Jac_\Hc(\F_q)\). 
Solving the DLP in the former group appears to be much more difficult 
than in the latter when the groups are roughly the same size, especially
for larger \(g\).

Indeed, Pollard's Rho algorithm is the best known
algorithm to solve DLP instances on a generic elliptic curve defined over a characteristic-two field of the form $\F_{q^\ell}$. 
This algorithm has an exponential computational complexity of 
$(1 + o(1)) O\left(\frac{1}{2}\sqrt{\pi\cdot q^\ell}\right)$~\cite{Galbraith2016,JoC_Oliveira}. 
On the other hand, using an index-calculus strategy one can solve the DLP on the Jacobian of a curve $\Hc$ over $\F_q$ 
with a subexponential complexity of $L_{q^g}\big[\frac{1}{2}, \sqrt{2} + o(1)\big]$ 
(as \(q\) and \(g\) tend to infinity).%
\footnote{%
    Recall that
    \(
        L_X[\alpha,c] 
        = 
        \operatorname{exp}\left( (c + o(1)) (\log X)^{\alpha} (\log \log X)^{1-\alpha}
        \right)
    \)
    for $0 < \alpha < 1$ and $c > 0$.
}

\paragraph{Weil descent}
The Weil descent attack was introduced by Frey in 1998 as a
means of transferring DLP instances
from an elliptic curve \(\Ec\) defined over an extension field  \(\F_{q^\ell}\)
to the Jacobian of a higher-genus curve \(\Hc \) defined over the subfield \(\F_q\)~\cite{Frey98}. 
This transfer becomes useful if the DLP in the Jacobian of the curve
\(\Hc / \F_q\) is easier than the DLP on \(\Ec /\F_{q^\ell} \), a situation that usually happens
if the genus $g$ of $\Hc$ is neither too large,
nor too small (i.e., \(g \geq \ell\) and \(g \approx \ell\)).

Frey's initial construction was refined by Galbraith and Smart in~\cite{GS99}. 
Gaudry, Hess, and Smart gave an efficient version of the Weil descent technique (GHS)
applied to curves defined over binary extension fields~\cite{GHS02a}.
Galbraith, Hess, and Smart extended this attack
to a larger class of curves by transferring the DLP to an isogenous
elliptic curve vulnerable to the GHS method~\cite{GHS02b},
and
Hess generalized the GHS Weil descent attack
from hyperelliptic \(\Hc/\F_q\)
to possibly non-hyperelliptic \(\Cc/\F_q\)~\cite{Hess03, Hess04}.

\paragraph{Our contributions}  
As explained above, Weil descent allows us to transfer DLP computations
from an elliptic curve \(\Ec / \F_{q^{\ell}}\) into the Jacobian of a genus-\(g\) curve \(\Hc/\F_q\). 
In this paper, we make three main contributions:
\begin{enumerate} \setcounter{enumi}{-1}
    \item 
        We show paper that if \(\Ec\) has a GLS endomorphism,
        then this induces an efficiently-computable endomorphism of \(\Jac_\Hc\).
        We give an explicit description of this endomorphism
        in~\S\ref{JAC:end}.
    \item We show that if \(\Jac_\Hc\) has an efficiently endomorphism
        with an eigenvalue of order $n$ on \(\Jac_\Hc(\F_q)\),
        then the relation generation stage of the index-calculus
        algorithm for solving the DLP in $\Jac_{\Hc}(\F_q)$
        can be accelerated by a factor of $n$.
        This in turn implies that the size of the factor base is
        reduced by a factor of $n$,
        which accelerates the linear algebra phase by a factor of $n^2$.
        We present an algorithmic analysis of the 
        expected speedup for discrete logarithm computations in  $\Jac_{\Hc}(\F_q)$
        in~\S\ref{sec:speedingup}.
    \item
        To illustrate our techniques,
        we present a concrete 115-bit discrete logarithm computation attacking a weak 
        GLS elliptic curve defined over the field $\F_{2^{5\cdot 31}}$ 
        (see~\S\ref{sec:toy} for a full description of the problem instance). 
        In our experiments,
        we observed a factor-$5$ speedup for the index-calculus 
        computation over $\Jac_\Hc(\F_{q})$. 
        In total, our proof-of-concept implementation 
        in the Magma computational algebra system~\cite{Magma}
        computed the
        discrete logarithm in just $1,035$ CPU days,
        which is significantly less than a discrete logarithm computation reported 
        for the same problem in~\cite{VJS14} (cf. Table~\ref{tb:exp:comp}). 
\end{enumerate}
To the best of our knowledge, the first two observations have not been previously reported in the literature.

\paragraph{Previous work}
In 2001, Menezes and Qu showed that the GHS attack cannot be applied
efficiently over binary fields with prime extension degree $n$ in the cryptographically interesting range 
\(n \in \{160,\ldots,600\}\)~\cite{MQ01}. Moreover, Jacobson, Menezes, and Stein studied the GHS Weil descent 
attack on elliptic curves defined over \(\F_{q^{31}}\) with \(q = 2^5\) \cite{JMS01}.
Maurer, Menezes, and Teske analysed the feasibility of the GHS attack 
on elliptic curves over binary extension fields with composite extension
degree $n$ in the interval \(n \in \{100,\ldots,600\}\)~\cite{MMT01}.
In 2009, Hankerson, Karabina and Menezes showed that 
binary Galbraith--Lin--Scott (GLS) elliptic curves (see~\S\ref{sec:bg}) defined over \(\F_{2^{2\ell}}\) are secure against the 
(generalized) GHS attack when \(\ell\) is a prime in \(\{80,\ldots, 256\} \setminus \{127\}\) \cite{HKM09}.
Finally, Chi and Oliveira presented an efficient algorithm to determine if a given GLS 
elliptic curve is vulnerable to the GHS attack~\cite{CO15}. 
In~\cite{VJS14}, Velichka \emph{et al.} presented an explicit computation of a discrete logarithm problem using
the Weil descent attack on a hyperelliptic genus-$32$ curve over $\F_{2^5}$.

Recently, but tangentially, Galbraith, Granger, Merz, and Petit~\cite{DBLP:journals/iacr/GalbraithGMP20}
showed how DLP computations on Koblitz curves can be sped up using
carefully-chosen factor bases, taking
advantage of the Frobenius endomorphism acting on these curves. Their techniques resemble the ones presented in 
this paper, since we reduce the factor base under endomorphism orbits defined on Jacobians of hyperelliptic 
curves. We believe that our factor base reduction can be easily adapted to the elliptic-curve setting 
from~\cite{DBLP:journals/iacr/GalbraithGMP20}, but this time applied to GLS curves.

\paragraph{Organization}  
This paper is structured as follows. 
We (briefly) provide mathematical background on 
hyperelliptic curves and a general description of the (g)GHS Weil descent attack
in~\S\ref{sec:bg}.
Generalized GLS binary curves and their endomorphisms are described in~\S\ref{sec:gls}.
In~\S\ref{sec:endom}, we present
a concrete formulation of the GLS endomorphism induced on the Weil restriction.
This is followed 
in~\S\ref{sec:new_endom} by a concrete definition of the GLS endomorphism on \(\Jac_\Hc(\F_q)\), which is the
main result of this paper, together with a detailed discussion of the
discrete logarithm computation 
in \(\Jac_\Hc(\F_q)\) by means of a standard index-calculus procedure. It is shown that the GLS endomorphism 
provides a factor-$n$ acceleration, in theory and in practice.
Concluding remarks are made in~\S\ref{sec:concl}.

\section{Mathematical background}\label{sec:bg}

We begin with some basic definitions and properties of [hyper]elliptic
curves, and a general description 
of the (g)GHS Weil descent attack. For more in-depth details, the interested reader is referred 
to~\cite{alfredhyper,ehcc,Galbraith12,Washington}.

\subsection{Binary GLS curves}
Let $q = 2^n$. 
A binary elliptic curve is given 
by the Weierstrass equation
\[
    \Ec / \F_{q^\ell}\colon y^2+xy = x^3 + ax^2 + b
    \,.
\]
The set of affine solutions $(x, y) \in \F_{q^\ell} \times \F_{q^\ell}$,
together with a point at infinity denoted by $\mathcal{O}$,
form an abelian group denoted by $\Ec(\F_{q^\ell})$. 
A careful selection of the constants $a, b$, 
yields a group order
$\#\Ec(\F_{q^\ell}) = c\cdot r$
where $r$ is a large prime, and $c$ a small cofactor.
Let $\langle \Pt\rangle$ be the order-$r$ subgroup of $\Ec(\F_{q^\ell})$.
Given an integer \(0 < k < r\), 
the elliptic curve scalar multiplication operation computes the multiple $\Qt=[k]\Pt$, 
corresponding to the sum of $k$ copies of $\Pt$.

GLS curves, introduced in~\cite{galbraith11},
are cryptographically interesting 
because they come equipped with an efficiently computable endomorphism $\psi$,
which can be used in the Gallant--Lambert--Vanstone (GLV) scalar
multiplication technique of~\cite{GLV01}.
This splits
the computation of $\Qt = [k]\Pt$ 
into two half-sized scalar multiplications such that
$$\Qt = [k]\Pt = [k_1]\Pt + [k_2]\psi(\Pt)\,,$$ 
which can be computed using a two-dimensional multiscalar multiplication
algorithm.
The authors of~\cite{HKM09} reported a family of binary GLS curves
over quadratic extensions $\mathbb{F}_{q^{2}}$ 
with almost-prime group orders of the form $\#\Ec_{a,b}(\F_{q^{2}}) = 2r$,
where $r$ is a $(2n-1)$-bit prime. The software and hardware implementations of constant-time 
variable-base-point elliptic curve scalar multiplication using binary GLS curves rank among the 
fastest at the 128-bit security level~\cite{AyMORS18,oliveira2016,oliveira2014_2}.

\subsection{Basic definitions and properties of hyperelliptic curves}
Let \(q = 2^n\) and let \(\ell > 1\) be an integer prime to \(n\).
Throughout this paper, 
$\Ec/\F_{q^\ell}$ is an elliptic curve
defined by
\begin{align} 
    \label{Eq:Ec}
    \Ec / \F_{q^\ell} \colon y^2 + x\cdot y = x^3 + a\cdot x^2 + b
    \quad\text{with}\quad 
    a \in \F_{q^\ell}
    \quad\text{and}\quad 
    b \not= 0\in \F_{q^\ell}
    \,,
\end{align}
while $\Hc / \F_{q^\ell}$ 
is a genus-\(g\) hyperelliptic curve 
defined by
\[
    \Hc / \F_{q^\ell} \colon y^2 + h(x)\cdot y = f(x),
\]
where $f, g \in \F_{q^\ell}[x]$ satisfy $\deg f = 2g + 1$ and $\deg h \leq g$.

The set of $\F_{q^\ell}$-rational points of $\Hc / \F_{q^\ell}$ is
\[
    \Hc \left( \F_{q^\ell} \right) = \left\{ (x,y) \in \F_{q^\ell}\times\F_{q^\ell} 
    \colon y^2 + h(x)\cdot y = f(x) \right\} \cup \{ \mathcal{O} \}
    \,,
\]
where $\mathcal{O}$ is the point at infinity. The opposite of any point 
$\Pt = (x,y) \in \Hc(\F_{q^\ell}) \setminus \{\mathcal{O}\}$ is defined as 
$\overline{\Pt} = (x, y + h(x))$. 
If \(g = 1\) then $\Hc$ is an elliptic curve, 
and \(\Hc(\F_{q^\ell})\) has a group law given by 
the usual chord-and-tangent rules. 
However, these rules are not well-defined when $g > 1$.
Instead, when $g > 1$
we work with the Jacobian $\Jac_\Hc$ of 
$\Hc$.
The group of points $\Jac_\Hc(\F_{q^\ell})$
can be defined in terms of the group of divisors of $\Hc / \F_{q^\ell}$.
A divisor $\Dv$ is a formal sum of points on the curve, i.e., 
$\Dv = \sum_{\Pt_i \in \Hc(\F_{q^\ell})}c_i(\Pt_i)$ where $c_i = 0$
for all but finitely many points $\Pt_i \in \Hc(\F_{q^\ell})$.
The degree of $\Dv$ is $\deg \Dv := \sum c_i$. 
Every nonzero rational function on \(\Hc\)
has an associated principal divisor.
In the language of divisors,
$\Jac_\Hc(\F_{q^\ell}) =
\Div^0_{\Hc}(\F_{q^\ell})/\Prin_{\Hc}(\F_{q^\ell})$,
where $\Div^0_{\Hc}(\F_{q^\ell})$ and $\Prin_{\Hc}(\F_{q^\ell})$ denote the 
groups of degree-zero and principal divisors on $\Hc$, respectively.

Algorithmically, it is more convenient to use the Mumford representation
for elements of $\Jac_\Hc(\F_{q^\ell})$.
Each divisor (class) is represented a pair of polynomials $u,v \in \F_{q^\ell}[x]$
such that \(u\) is monic with \(\deg u \leq g\),
and \(v\) satisfies \(\deg v < \deg u\)
and \(u \mid (v^2 + v h - f)\).
If \(\Dv = \sum_{i=1}^g c_i(P_i) - g(\mathcal{O})\),
then 
\(x(P_i)\) is a root of \(u\) with multiplicity \(c_i\),
and $v(x(P_i)) = y(P_i)$.
The divisor corresponding to the pair $(u,v)$ is denoted $\dve(u,v)$.
The group law on divisors in the Mumford representation can be computed
using Cantor's algorithm~\cite{Washington}.

Mumford's representation allows us to define notions of irreducibility 
and smoothness for divisors:
\begin{enumerate}
    \item
        $\dve(u,v)$ is irreducible if $u$ is irreducible, and 
    \item
        $\dve(u,v)$ is $s$-smooth if $u$ is $s$-smooth.
\end{enumerate}
An important and useful fact is that if $u = \prod_i u_i$ then $\dve(u,v) = \sum_i \dve(u_i, v \mod u_i)$. 

By the Riemann--Roch theorem,
every divisor class in \(\Jac_\Hc\)
can be represented by a sum of divisors in the form
\((\Pt) - (\mathcal{O})\)
with \(\Pt \in \Hc(\overline{\F}_{q^\ell})\).
If \(\Pt = \left( x_\Pt , y_\Pt\right)\),
then
\((\Pt) - (\mathcal{O}) = \dve\big(x + x_\Pt, y_\Pt\big)\).
Consequently, any divisor \(\dve(u,v) \in \Jac_{\Hc}(\F_{q^\ell})\) can be written as \(\sum_i c_i\cdot \dve\big(x + x_{\Pt_i}, y_{\Pt_i}\big)\), where \((x_{\Pt_i},y_{\Pt_i})\in \Hc(\overline{\F}_{q^\ell})\), \(u~=~\prod_i {(x + x_{\Pt_i})}^{c_i}\), and \(v(x_{\Pt_i}) = y_{\Pt_i}\).

\begin{remark}
    The Jacobian of any elliptic curve \(\Ec / \F_{q^\ell}\) is 
    isomorphic to its group of rational points, i.e., $\Jac_\Ec(\F_{q^\ell}) \cong \Ec(\F_{q^\ell})$.
\end{remark}

\subsection{Computing discrete logarithms on hyperelliptic curves}
\label{sec:index-calculus}

As we mentioned in the introduction, 
the (g)GHS Weil descent technique permits to reduce the DLP 
in $\Ec (\F_{q^\ell})$ into the \(\Jac_{\Hc}(\F_q)\), where  \(\Hc / \F_q\) is  a hyperelliptic genus-\(g\) 
curve defined over $\F_q$~\cite{GHS02a, GHS02b, Hess03, Hess04}. 
Suppose, then,
that we want to solve a DLP instance $\Dv' = \lambda \Dv$ in
$\Jac_\Hc(\F_{q^\ell})$,
where \(\Dv\) and $\Dv' \in \langle \Dv \rangle$ have prime order $r$.
The most efficient method for solving the DLP on $\Jac_\Hc(\F_{q^\ell})$
is an index-calculus approach, consisting of the following steps.

Fix a smoothness bound \(s\),
and choose a small positive integer \(\epsilon\).
Let \(F(s)\) be  the number of irreducible 
divisors \(\dve(u,v) \in \Jac_H(\F_q)\) with \(\deg u \leq s\);
these divisors form the factor base.
We need to generate $F(s) + \epsilon$ relations of the form
$\alpha_i \Dv + \beta_i\Dv'~=~\sum_{j=1}^{F(s)} m_{i,j}\Dv_j$,
with the \(\Dv_j\) in the factor base,
in order to construct three matrices 
\(\alpha = (\alpha_i)^\mathsf{T}\), 
\(\beta = (\beta_i)^\mathsf{T}\),
and  
\(M = (m_{i,j})\) with coefficients in \(\Z/r\Z\). Once 
that this task is completed, we compute an element \(\gamma\) 
of the kernel of \(M^\mathsf{T}\);
then
\(\left(\gamma^\mathsf{T} \alpha\right)\Dv + \left(\gamma^\mathsf{T} \beta\right)\Dv' = 0\). 
If \(\gamma^\mathsf{T} \beta = 0\), then we must repeat the whole
procedure (or at least try a different \(\gamma\)); otherwise, 
the discrete logarithm of $\Dv'$ with respect to $\Dv$ is 
$\lambda = -(\gamma^\mathsf{T} \alpha)/(\gamma^\mathsf{T} \beta)$.

When the genus of  the curve produced by the (g)GHS Weil descent attack is large with respect to the finite 
field size, the most efficient choice for the DLP on higher-genus
hyperelliptic curves is the Enge-Gaudry 
algorithm \cite{Gaudry00,EG02},
with a
subexponential running-time complexity of 
\[ L_{q^g}\left[\frac{1}{2}, \sqrt{2} + o(1)\right] = \mbox{exp}{\left(\sqrt{2} + o(1)\right)\sqrt{\log q^g}\sqrt{\log \log q^g}}.\]

\label{gGHS} We say that the elliptic curve $\Ec / \F_{q^\ell}$ is vulnerable (or weak) against the 
(g)GHS Weil descent attack if the computational cost of the
Enge-Gaudry algorithm on the hyperelliptic curve constructed by the GHS
attack is less than that of Pollard's rho algorithm.

In the concrete discrete logarithm computation of~\S\ref{sec:toy}, (g)GHS Weil descent produces a hyperelliptic
genus-32 curve $\Hc / \F_{q}$, with $q = 2^5$. In other words, for our discrete logarithm computation we work with 
the case $g = q$.

\begin{remark}
    For curves of small genus \(g\geq 3\), the algorithm of Gaudry,
    Thom\'e, Th\'eriault, and Diem~\cite{GTTD07} is the most efficient 
    choice for solving DLPs. For genus-\(2\) curves, 
    Pollard's Rho algorithm is more efficient.
\end{remark}

\subsection{Costs of the index-calculus based algorithm}\label{subsec:cost:index}
The two main steps of the index-calculus approach are the search for
$s$-smooth divisors, and the computation of a kernel element, which is handled as a linear algebra problem.
For the first task, 
one can approximate the cost of finding \(s\)-smooth divisors search as
follows (for more details see~\cite{JMS01}):
If $A_{s'}$ is the number of irreducible divisors \(\dve(u,v) \in \Jac_H(\F_q)\) with \(\deg u = s'\), then
\[
    A_{s'} \approx \frac{1}{2} \cdot \frac{1}{s'} \sum_{d \mid s'} \mu
    \Big( \frac{s'}{d}\Big) q^d 
    \,,
\]
where $\mu$ denotes the M\"{o}bius function, i.e., 
\( \mu (n) = (-1)^k\) if \(n\) is squarefree with \(k\) different prime
factors, and
\(0\) if \(n\) is not squarefree.
Consequently, \(F(s) \approx \sum_{i=1}^{s} A_i\).
On the other hand, the number of $s$-smooth divisors $\dve(u,v) \in \Jac_{\Hc}(\F_q)$ 
with \(\deg u \leq g\) is 
\[
    M(g,s) = \sum_{i = 1}^{g}\left( \left[x^i\right] \prod_{s' = 1}^s {\left( \frac{1 + x^{s'}}{1 - x^{s'}}\right)}^{A_{s'}} \right)
    \,,
\]
where $[.]$ denotes the coefficient operator.
When $A_{s'}$ is known, $M(g,s)$ 
can be computed by finding the first $(g+1)$ terms of the Taylor expansion of 
$\prod_{s' = 1}^s {\left( \frac{1 + x^{s'}}{1 - x^{s'}}\right)}^{A_{s'}}$ around 
$x = 0$, and summing the coefficients of $x, x^2, \ldots, x^g$. 

The expected 
number of random-walk steps before encountering an $s$-smooth divisor is
therefore
\[
    E(s) = \frac{\# \Jac_{\Hc}(\F_q)}{M(g,s)} \approx \frac{q^g}{M(g,s)}
    \,,
\]
and the expected number of steps before 
$F(s) + \epsilon$ relations are generated is
\[
    T(s) = \left(F(s) + \epsilon\right) E(s)
    \,.
\]

For the linear algebra task, 
Magma uses Lanczos' algorithm, with approximate running time 
$L(s) \approx d\cdot(F(s) + \epsilon)^2$
where \(d\) denotes
the per-row density of the matrix $M$.
In fact, it can be shown that \(d \leq g\).

\section{The GLS endomorphism}\label{sec:gls}
Let $\Ec$ be an elliptic curve over $\F_{q^\ell}$,
defined by Equation~\eqref{Eq:Ec} 
with \(a\in\F_q\subset\F_{q^\ell}\) 
and \(b\in\F_{2^\ell}\subset\F_{q^\ell}\). 
For each integer \(i \geq 0\),
we define an elliptic curve
\[
    \Ec_i / \F_{q^\ell} 
    \colon 
    y^2 + xy = x^3 + a^{2^i}x^2 + b^{2^i}
    \,.
\]
The curves 
\(\Ec = \Ec_0, \Ec_1, \ldots, \Ec_{n\cdot\ell-1}, \Ec_{n\cdot\ell} = \Ec\)
are connected by a cycle of \(2\)-power Frobenius maps
$\Ec_i / \F_{q^\ell} \to \Ec_{i+1} / \F_{q^\ell}$ 
mapping $(x,y) \mapsto ( x^2, y^2)$.
Abusing notation, 
we will write \(\pi\) for each of these maps
and \(\pi^k\) for the composition of any \(k\) successive ones.
Since \(b\) is in \(\F_{2^\ell}\),
the curve $\Ec_\ell / \F_{q^\ell}$ 
is isomorphic to $\Ec / \F_{q^\ell}$; 
the isomorphism 
is 
\begin{align*} 
    \phi \colon \Ec_\ell / \F_{q^\ell} 
    &
    \longrightarrow \Ec / \F_{q^\ell} 
    \\
    (x,y) 
    &
    \longmapsto \left(x,y + \delta x \right),
\end{align*}
where $\delta^2 + \delta = a + a^{2^\ell}$. 
If $n\cdot\ell$ is odd,
then $\delta \in \F_{q^\ell}\setminus\F_{2^\ell} $, 
so the isomorphism \(\phi\) is defined over \(\F_{q^\ell}\),
and in particular
$\delta = \sum_{j = 0}^{ \frac{n\cdot\ell - 1}{2}}{\left(a + a^{2^\ell}\right)}^{2^{2j}}$. 

Composing the \(2^\ell\)-power Frobenius \(\pi^\ell \colon \Ec \to \Ec_\ell\) 
with the isomorphism \(\phi \colon \Ec_\ell \to \Ec\),
we obtain a generalized Galbraith--Lin--Scott (GLS) endomorphism
\[
    \psi \coloneqq \phi\circ \pi^\ell \colon (x,y) 
    \longmapsto
    \big(x^{2^\ell},y^{2^\ell}+\delta x^{2^\ell}\big)
    \in \mathrm{End}(\Ec).
\]
The endomorphism \(\psi\) is defined over $\F_{q^\ell}$ and satisfies
\(\psi^n = \pm\pi^{n\ell}\);
in particular, \(\psi^n\) acts as \([1]\) or \([-1]\)
on points of~\(\Ec(\F_{q^\ell})\).

Endomorphisms such as \(\psi\) are cryptographically interesting  because they can be used to accelerate scalar multiplication on 
\(\Ec\), by applying the technique of Gallant, Lambert, and Vanstone (for more details, see \cite{GLV01}). 
In the sequel, we will show that these 
endomorphisms can also be used to improve the efficiency of the Gaudry--Hess--Smart Weil descent attack on weak curves of this kind.

\section{Extending the GLS endomorphism}\label{sec:endom}
From now on, we fix an element \(w\) of \(\F_{q^\ell}\)
such that
$w + w^2 + \cdots + w^{2^{\ell - 1}} = 1$ 
and
\[
    \F_{q^\ell} 
    = 
    \F_q(w) 
    = 
    {\big\langle w,w^2,w^4,\ldots, w^{2^{\ell-1}}\big\rangle}_{\F_q}
    ;
\] 
that is, \(\{w,w^2,w^4,\ldots,w^{2^{\ell-1}}\}\) 
is a normal basis for \(\F_{q^\ell}\) over \(\F_q\).

Recall that 
the Weil restriction 
\[
    \Ac_i / \F_q 
    \coloneqq 
    \W_{\F_q}^{\F_{q^\ell}}\left( \Ec_i \right)
\]
of $\Ec_i$ from $\F_{q^\ell}$ to~\(\F_{q}\)
is an \(\ell\)-dimensional abelian variety
over \(\F_{q}\), 
and that there is an isomorphism of groups
\(\Ac_i(\F_{q})\cong \Ec_i(\F_{q^\ell})\).\footnote{%
    More generally, for any algebra \(K\) over \(\F_q\),
    there is an isomorphism 
    between \(\Ec_i(\F_{q^\ell}\otimes_{\F_q}K)\)
    and \(\Ac_i(K)\);
    in fact, \(\Ac_i\) is the group scheme
    realizing the functor 
    \(K \mapsto \Ec_i(\F_{q^\ell}\otimes_{\F_q}K)\).
}
The various isogenies and endomorphisms of \(\Ec_i\)
induce isogenies and endomorphisms of \(\Ac_i\).

We will use the following explicit affine model for~\( \Ac_i \).
Consider the polynomial ring
\(R = \F_q[x_0,x_1,\ldots,x_{\ell-1},y_0,\ldots,y_{\ell-1}]\),
and set 
\[
    X = \sum_{j=0}^{\ell-1}x_jw^{2^j}
    \qquad
    \text{and}
    \qquad
    Y = \sum_{j=0}^{\ell-1}y_jw^{2^j}
\]
in \(R\otimes\F_{q^\ell}\).
Expanding the defining equation of \(\Ec_i\)
in the variables \(X\) and \(Y\),
there exist 
\(W_0,\ldots,W_{\ell-1}\) in \(R\)
such that 
\(Y^2 + XY - (X^3 - (a^{2^i})X^2 - b) = \sum_{j=0}W_jw^{2^j}\)
in \(R\otimes\F_{q^\ell}\).
The affine scheme \(\mathrm{Spec}(R/(W_0,\ldots,W_{\ell-1}))\)
is then \(\F_q\)-isomorphic to 
an open affine subset of \(\Ac_i\).
By construction,
we have a bijection of sets
\begin{align*}
    \iota \colon \Ec_i(\F_{q^\ell})
    &
    \longrightarrow \Ac_i(\F_q)
    \\
    (x, y) 
    &
    \longmapsto 
    (x_0,\ldots,x_{n-1},y_0,\ldots,y_{n-1}),
\end{align*}
where $x = \sum_{j = 0}^{\ell-1} x_j w^{2^j}$ 
and $y = \sum_{j = 0}^{\ell-1} y_j w^{2^j}$. In fact, 
\(\iota\) is an isomorphism of groups.

We want to make the isogenies and endomorphisms of \(\Ac_i\)
corresponding to \(\pi\), \(\phi\), and \(\psi\)
completely explicit with respect to this affine model of \(\Ac_i\).
First, observe that if \(X = \sum_{j=0}^{\ell-1} x_jw^{2^j}\),
then \(X^2 = \sum_{j=0}^{\ell-1} x_j^2w^{2^{j+1}}\),
so the \(2\)-powering Frobenius isogeny \(\pi: \Ec_i \to \Ec_{i+1}\)
corresponds to an isogeny \(\Pi: \Ac_i \to \Ac_{i+1}\)
that squares and cyclically permutes the coordinates:
\[
    \Pi:
    (x_0,\ldots,x_{\ell-1},y_0,\ldots,y_{\ell-1})
    \longmapsto
    (x_{\ell-1}^2,x_0^2,\ldots,x_{\ell-2}^2,
    y_{\ell-1}^2,y_0^2,\ldots,y_{\ell-2}^2)
    .
\]

The isomorphism $\phi: \Ec_\ell \to \Ec$ maps \((X,Y)\) to \((X,Y+\delta X)\),
and \(\delta\) is in \(\F_q\) because \(a\) is,
so \(\phi\)
corresponds to an isomorphism \(\Phi: \Ac_\ell \to \Ac\) defined by
\[
    \Phi:
    (x_0,\ldots,x_{\ell - 1},y_0,\ldots,y_{\ell - 1}) 
    \longmapsto 
    \left(
        x_0,\ldots,x_{\ell-1},y_0 + \delta x_0,
        \ldots, 
        y_{\ell-1} + \delta x_{\ell-1} 
    \right)
    .
\]
As with \(\pi^\ell\) and \(\phi\) on the elliptic curves,
composing \(\Pi^\ell: \Ac \to \Ac_\ell\)
with \(\Phi: \Ac_\ell \to \Ac\)
yields an endomorphism \(\Psi\) of \(\Ac\),
defined (over \(\F_q\)) by
\begin{align*} 
    \Psi:
    (x_0,\ldots,x_{\ell - 1},y_0,\ldots,y_{\ell - 1}) 
    \longmapsto 
    \left(
        x_0^{2^\ell},\ldots,x_{\ell-1}^{2^\ell},
        y_0^{2^\ell} + \delta x_0^{2^\ell}, 
        \ldots, 
        y_{\ell-1}^{2^\ell} + \delta x_{\ell-1}^{2^\ell}
    \right).
\end{align*} 

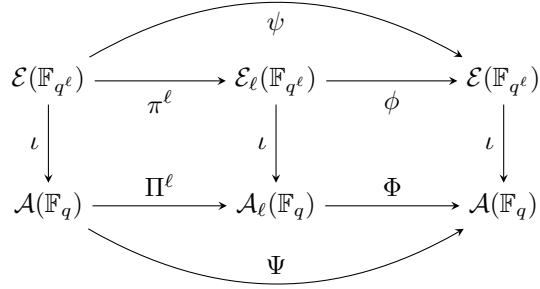
\begin{figure}[!hbt]
    \centering
    \begin{tikzpicture}
        \matrix (m) [matrix of math nodes,row sep=3em,column sep=5em,minimum width=2em] {
            \Ec(\F_{q^\ell})
            & 
            \Ec_\ell(\F_{q^\ell})
            & 
            \Ec(\F_{q^\ell})
            \\
            \Ac(\F_q)
            & 
            \Ac_\ell(\F_q)
            & 
            \Ac(\F_q)
            \\
        };
        \path[-stealth]
            (m-1-1) edge node [below] {$\pi^\ell$} (m-1-2)
	        edge node [left] {$\iota$} (m-2-1)
	        edge [bend left] node [below] {$\psi$} (m-1-3)
            (m-1-2) edge node [below] {$\phi$} (m-1-3)
	        edge node [left] {$\iota$} (m-2-2)
            (m-2-1) edge node [above] {$\Pi^\ell$} (m-2-2)
            edge [bend right] node [above] {$\Psi$} (m-2-3)
            (m-2-2) edge node [above] {$\Phi$} (m-2-3)
            (m-1-3) edge node [left] {$\iota$} (m-2-3);
    \end{tikzpicture}
    \caption{Endomorphism diagram}
    \label{fig:diag}
\end{figure}

On groups of points we have
\(\Pi = \iota\circ\pi\circ\iota^{-1}\),
$\Phi = \iota \circ \phi \circ \iota^{-1}$,
and
$\Psi = \iota \circ \psi \circ \iota^{-1}$.
The relationships between all of these various maps
are summarized in Figure~\ref{fig:diag}.

We note that if \(\Grp\) is a cyclic subgroup of
\(\Ec(\F_{q^\ell})\)
of order \(r\),
and \(\psi\) acts on \(\Grp\) 
as multiplication by some eigenvalue \(\lambda\pmod{r}\),
then 
\(\Psi\) must act on \(\iota(\Grp) \subseteq \Ac(\F_q)\) 
as multiplication by exactly the same eigenvalue \(\lambda\).

\section{Combining the GLS and GHS techniques}\label{sec:new_endom}
The generalized GHS (gGHS) Weil descent technique 
constructs a genus-$g$ algebraic curve $\Cc / \F_q$ (not necessary hyperelliptic)
by computing the Weil restriction $\Ac / \F_q$ of $\Ec / \F_{q^\ell}$, intersecting $\Ac / \F_q$  
with $(\ell - 1)$-dimensional hyperplanes to obtain a subvariety $\Ac' / \F_q$ of $\Ac / \F_q$,  
and finding an irreducible component $\Cc / \F_q$ of $\Ac' / \F_q$ (for more details see 
\cite{GHS02a, GHS02b, Hess03, Hess04}).

Let us intersect $\Ac / \F_q$
with the hyperplanes $x_0 = x_1 =\cdots=x_{\ell-1} = x \in \F_q$.
With
\(a\in\F_q\) and \(b\in\F_{2^\ell}\), and using the linear independence property 
of a normal basis \(\{w,w^2,w^4,\ldots,w^{2^{\ell-1}}\}\),
we obtain a subvariety $\Ac' / \F_q$ of $\Ac / \F_q$
defined by
\begin{align*} 
    \Ac' / \F_q \colon \left\{
    \begin{array}{rll}
        x^3 + a\cdot x^2 + x\cdot y_0 + y_{\ell - 1}^2 + b_0 &= &0 \\ 
        x^3 + a\cdot x^2 + x\cdot y_1 + y_0^2 + b_1 &= &0 \\
        &\vdots \\
        x^3 + a\cdot x^2 + x\cdot y_{\ell - 1} + y_{\ell - 2}^2 + b_{\ell - 1} &= &0 \\
    \end{array}
    \right.
\end{align*} 
where \(b = \sum_{i=0}^{\ell-1}b_iw^{2^i}\) and each \(b_i\) is in \(\F_2\). Thus, 
if \(\Ac'_\ell / \F_q\) is the variety determined by Equation~(\ref{eq:ConjACc}), then 
\(\Phi\) induces an endomorphism of \(\Ac'_\ell(\F_q)\).
\begin{align} 
\Ac'_\ell / \F_q \colon \left\{
\begin{array}{rll}
x^3 + a^{2^\ell}\cdot x^2 + x\cdot y_0 + y_{\ell - 1}^2 + b_0 &= &0 \\ 
x^3 + a^{2^\ell}\cdot x^2 + x\cdot y_1 + y_0^2 + b_1 &= &0 \\
&\vdots \\
x^3 + a^{2^\ell}\cdot x^2 + x\cdot y_{\ell - 1} + y_{\ell - 2}^2 + b_{\ell - 1} &= &0 \\
\end{array}
\right.
\label{eq:ConjACc}
\end{align} 

\subsection{New endomorphism on the hyperelliptic curve}

Let \(\Hc / \F_q \colon y^2 + h(x)\cdot y = f(x)\) be a genus-\(g\) hyperelliptic curve 
that is an irreducible component of \(\Ac' / \F_q\). 
Writing \(h(x) = \sum_{i=0}^g h_i x^i\) 
and \(f(x)~=~\sum_{i=0}^{2g+1} f_i x^i\), the corresponding hyperelliptic irreducible 
component \(\Hc_\ell / \F_q\) of \(\Ac'_{\ell} / \F_q\) is 
\[
    \Hc_\ell / \F_q \colon y^2 + (^\sigma\! h)(x)\cdot y = (^\sigma\! f)(x)
\]
where \((^\sigma\! h)(x) = \sum_{i=0}^g \sigma(h_i) \cdot x^i\), \((^\sigma\! f)(x) = \sum_{i=0}^{2g+1} \sigma(f_i) \cdot x^i\), 
and \(\sigma(x) = x^{2^\ell}\) for all \(x\in\F_q\). 
Therefore, the maps \(\Pi^\ell \colon \Hc /\F_q \to \Hc_\ell /\F_q \) and \(\Phi \colon \Hc_\ell /\F_q \to \Hc /\F_q \) 
are defined by
\begin{align*} 
    \Pi^\ell \colon (x,y) \longmapsto \big(x^{2^\ell}, y^{2^\ell}\big)
    \quad \text{and}\quad 
    \Phi \colon (x,y) \longmapsto \big(\delta_1\cdot x + \delta_2, \delta_3\cdot y + t(x)\big)
\end{align*} 
for some \(\delta_1,\delta_2,\delta_3 \in \F_q\) and \(t(x) \in
\F_q[x]\) with \(\deg t(x) \leq g\) and \(\delta_1\neq 0\).\footnote{Any
isomorphism of hyperelliptic curves over a finite field is in the form of \(\Phi\) (for more details see~\cite[Section 10.2]{Galbraith12}).}
Consequently, $\Psi = \Phi \circ \Pi^\ell$ induces the following endomorphism:
\begin{align*} 
    \begin{split}
        \Psi^\ast \colon \Jac_{\Hc}\left( \overline{\F_{q}} \right) 
        &
        \longrightarrow \Jac_{\Hc}\left( \overline{\F_{q}} \right) 
        \\
        \sum_j c_j \left( \Pt_j \right) 
        &
        \longmapsto \sum_j c_j \left( \Psi (\Pt_j) \right).
    \end{split}
\end{align*} 
In Mumford's representation,
the divisor 
\(\dve (u,v) = \sum_j c_j\cdot \dve (x + x_{\Pt_j}, y_{\Pt_j})\) is mapped 
to \(\sum_j c_j\cdot \dve (x + x_{\Psi(\Pt)_j}, y_{\Psi(\Pt)_j})\), 
and therefore
\(\F_q\)-irreducible factors of \(u\) are mapped to 
irreducible factors of the same degree, i.e., $\Psi^\ast$ sends smooth 
divisors to smooth divisors. 

The curve \(\Hc / \F_q\) has genus \(g \ge \ell\), so its Jacobian 
\(\Jac_{\Hc}\) is \(g\)-dimensional. By the universal property of the 
Jacobian,
the \(\ell\)-dimensional \(\Ac\) is a quotient (and so an isogeny 
factor) of \(\Jac_{\Hc}\).\footnote{Universal property: 
let \(\kappa\colon \Hc \to \tilde{\Ac}\) be a morphism, where 
\(\tilde{\Ac}\) is an abelian variety. Let \(\Pt_0 \in \Hc(\overline{\F_q})\) 
be such that \(\kappa(\Pt_0) = 0 \), and consider the map 
\(\tilde{\kappa} \colon \Hc \to \Jac_\Hc\) given by \(\Pt \mapsto \big(\Pt\big) - \big(\Pt_0\big)\).
Then there is a unique homomorphism \(\psi \colon \Jac_\Hc \to \tilde{\Ac}\) of abelian 
varieties such that \(\kappa = \psi \circ \tilde{\kappa}\) (for more
details see~\cite[Section 10.5]{Galbraith12}).} 
Hence, $\Jac_{\Hc} \cong \Ac \times \mathcal{B}$ 
for some \((g-\ell)\)-dimensional abelian variety $\mathcal{B}$. The situation 
is illustrated by the diagram in Figure~\ref{fig:diag:last}.

\begin{figure}[!hbt]
    \centering
    \begin{tikzcd}
        & 
        \Jac_{\Hc}  \arrow{d} \arrow{r}{\text{$\Psi^\ast$}} 
        & 
        \Jac_{\Hc}  \arrow{d}
        \\
        \Hc  \arrow[loop left, "\Psi"] \arrow[r] \arrow[ru, hook] 
        & 
        \Ac \times \mathcal{B} \arrow{r}{\text{$\Psi$}} 
        & 
        \Ac \times \mathcal{B}
    \end{tikzcd}
    \caption{Endomorphism diagram for $\Hc / \F_q$}
    \label{fig:diag:last}
\end{figure}

If \(\Grp\) is a cyclic subgroup of \(\Ec(\F_{2^{\ell\cdot n}})\)
fixed by \(\psi\),
then \(\psi\) acts on \(\Grp\) as multiplication by an eigenvalue \(\lambda\),
and so \(\Psi\) acts on \(\iota(\Grp)\) as multiplication by \(\lambda\). 
Hence, \({\left(\Psi^\ast\right)}^n = [1]\) or \({\left(\Psi^\ast\right)}^n = [-1]\), 
and therefore 
\(t(x) = \delta_4(^\sigma\! h)(\delta_5\cdot x)\) for some \(\delta_4, \delta_5 \in \F_q\). 
The morphism \({\Psi}^n \colon \Hc/\F_q \to \Hc/\F_q\) fixes the
$x$-coordinate, and
\[
    x_{{\Psi}^n(\Pt)} 
    = 
    \delta_1^{\left(\sum_{k=0}^{n-1}{{2^{k\cdot\ell}}}\right)}\cdot x_{\Pt}^{\left(2^{n\cdot\ell}\right)} 
    + 
    \sum_{i=0}^{n-1}\delta_1^{\left(\sum_{k=0}^{i-1}2^{k\cdot \ell}\right)}\cdot\delta_2^{\left(2^{i\cdot\ell}\right)}
    \,.
\]
But \((\ell,n) = 1\) and \(q=2^n\),
so
\(\delta_1^{\left(\sum_{k=0}^{n-1}{{2^{k\cdot\ell}}}\right)} = {(\delta_1)}^{2^n - 1} = {(\delta_1)}^{q - 1} = 1\),
while \(x_{\Pt}^{2^{n\cdot\ell}} = x_{\Pt}^{q^\ell} = x_{\Pt}\) 
and \(x_{{\Psi}^n(\Pt)} =  x_{\Pt} + \sum_{i=0}^{n-1}\delta_1^{\left(\sum_{k=0}^{i-1}2^{k\cdot \ell}\right)}\cdot\delta_2^{\left(2^{i\cdot\ell}\right)}\). 
Therefore, \(\sum_{i=0}^{n-1}\delta_1^{\left(\sum_{k=0}^{i-1}2^{k\cdot \ell}\right)}\cdot\delta_2^{\left(2^{i\cdot\ell}\right)} = 0\).
It follows that \(\delta_2 = 0\). 

\subsection{Explicit description of the new endomorphism}
\label{JAC:end}

Recall that for any point \(\Pt = \left( x_\Pt , y_\Pt\right) \in \Hc(\F_q)\)
its corresponding divisor \((\Pt) - (\mathcal{O})\) is equal to \(\dve\big(x + x_\Pt, y_\Pt\big)\), 
and any divisor \(\dve(u,v) \in \Jac_{\Hc}(\F_{q^\ell})\) can be written as 
\(\sum_i c_i\cdot \dve\big(x + x_{\Pt_i}, y_{\Pt_i}\big)\), where 
\((x_{\Pt_i},y_{\Pt_i})\in \Hc(\overline{\F}_{q^\ell})\), \(u = \prod_i {(x + x_{\Pt_i})}^{c_i}\), and \(v(x_{\Pt_i}) = y_{\Pt_i}\). 
The divisor \( \Psi^\ast\big((\Pt)\big) = \big( \Psi (\Pt)\big)\) is
therefore equal to 
\(\dve\big( x + \big(\delta_1x_\Pt^{2^\ell}\big), \delta_3y_\Pt^{2^\ell} + \delta_4(^\sigma\! h)\big(\delta_5\delta_1x_\Pt^{2^\ell}\big)\big)\), and any 
divisor \(\dve (u,v) \in \Jac_\Hc(\F_q)\) satisfies 
\(\dve (u,v) = \sum_{i=0}^{\deg u} \dve(x + x_i, v(u_i)) = \sum_{i=0}^{\deg u} \big((x_i, v(x_i))\big)\), 
where \(x_0\), \(x_1,\ldots,x_{\deg u} \in \overline{\F}_q\) are the roots of \(u\). 
Further,
\begin{align*} 
        \Psi^\ast \big(\dve(u,v)\big) &= \Psi^\ast \Bigg(\sum_{i=0}^{\deg u} \Big(\big(x_i, v(x_i)\big)\Big)\Bigg)
        = 
        \sum_{i=0}^{\deg u} \Psi^\ast\Big(\big(x_i, v(x_i)\big)\Big)
        \\
        &= 
        \sum_{i=0}^{\deg u} \dve\Big( x + \big(\delta_1{x_i}^{2^\ell}\big), \delta_3{\big(v(x_i)\big)}^{2^\ell} + \delta_4(^\sigma\! h)\big(\delta_5\delta_1{x_i}^{2^\ell}\big)\Big) 
        \\
        &= 
        \sum_{i=0}^{\deg u} \dve\Big(
            \delta_1\Big( \frac{x}{\delta_1} + {x_i}^{2^\ell}\Big), 
            \delta_3(^\sigma\! v)\big(x_i^{2^\ell}\big) + 
            \delta_4(^\sigma\! h)\big(\delta_5\delta_1{x_i}^{2^\ell}\big)\Big)
        \,.
\end{align*} 
We want to find polynomials \(u^\ast, v^\ast \in \F_q[x]\) such that 
\(\Psi^\ast \big(\dve(u,v)\big) = \dve(u^\ast, v^\ast)\), \(u^\ast\big(\delta_1{x_i}^{2^\ell}\big) = 0\), 
and \(v^\ast\big(\delta_1{x_i}^{2^\ell}\big) = \delta_3{\big(v(x_i)\big)}^{2^\ell} + \delta_4(^\sigma\! h)\big(\delta_5\delta_1{x_i}^{2^\ell}\big)\). 
In particular, \(u^\ast (x) = \delta_1^{\deg u}\cdot (^\sigma\! u)\big(\frac{x}{\delta_1}\Big) = \prod_{i=0}^{\deg u} \Big( x + \delta_1\cdot{x_i}^{2^\ell}\big)\),
and
\(v^\ast (x) = \delta_3(^\sigma\! v)\Big(\frac{x}{\delta_1}\Big) + \delta_4\big(^\sigma\!h\big)(\delta_5x)\).
Moreover,
\(\deg v^\ast < \deg u^\ast \leq g\) and 
\(u^\ast \mid \big({(v^\ast)}^2 + (v^\ast \cdot h) + f\big)\), so we can set 
\(v^\ast (x)\coloneqq \delta_3({}^\sigma v)\big(\frac{x}{\delta_1}\big)
+ \big(\delta_4\big({}^\sigma h\big)(\delta_5x) \big)\mod u^\ast(x)\).

The endomorphism \(\Psi^\ast\) must be well-defined in the sense that \(v^\ast\) should be 
the same if we reduce \(h\) modulo \(u\) from the beginning. This observation implies
\begin{align} 
    \delta_4({}^\sigma h)(\delta_5x)
    \equiv
    \delta_4\big({}^\sigma (h\bmod{u})\big)(\delta_5x)
    \pmod{u^\ast(x)}
    \,.
    \label{eq:hMOD}
\end{align} 

Write \(u = \sum_i u_ix^i\), 
\(v = \sum_i v_ix^i\), and \(h = \sum_i h_i x^i\).
If \((h\mod u) = \sum_i h'_i x^i\) and 
\(\delta_4({}^\sigma h)(\delta_5x)\mod u^\ast(x) = \sum_i h^\ast_{i}x^i\),
then we have
\(u^\ast = \delta_1^{\deg u}\sum_i ({u_i}^{2^\ell}/\delta_1^i) x^i\), 
\(\delta_3(^\sigma\! v)(x/\delta_1) = \delta_3\sum_i ({v_i}^{2^\ell}/\delta_1^i) x^i\), and 
\( \delta_4\big({}^\sigma(h\bmod u)\big)(\delta_5x) = \delta_4\sum_i {\big(h'_i\big)}^{2^\ell}\delta_5^i x^i\). 
In particular, Equation (\ref{eq:hMOD}) holds for any \(\dve(u,v) \in \Jac_\Hc(\F_q)\). 

Let us analyze the cases \(\deg u = g \) and \(\deg u \le g \) separately.

\paragraph{The case \(\deg u = g \)}
In this case, we can write
\begin{align} 
    h'_i = c_{h_i} + h_g\cdot u_i 
    \quad \text{and}\quad
    h^\ast_i= \delta_4\cdot\Big({h_i}^{2^\ell}\cdot\delta_5^i + \big({h_g}^{2^\ell}\cdot\delta_5^g\big)\cdot\delta_1^{g-i}\Big).
    \label{eq:delta5}
\end{align} 
Taking Equations~\eqref{eq:hMOD} and~\eqref{eq:delta5} together, for each \(i=0,\ldots, (g-1)\) we have
\[
    \delta_4\cdot{(h_i + h_g\cdot u_i)}^{2^\ell}\cdot \delta_5^i 
    =
    \delta_4\cdot\big({h_i}^{2^\ell}\cdot\delta_5^i + \big({h_g}^{2^\ell}\cdot\delta_5^g\big)\cdot\delta_1^{g-i}\big)
    \,.
    \label{eq:delta5:2}
\]
This equation is satisfied if and only if \(\delta_5^i = \delta_5^g\cdot\delta_1^{g-i}\) 
for each \(0 \le i < g\).
Moreover, \(\delta_5 = \frac{1}{\delta_1}\) and
\[
    \Psi^\ast \big(\dve(u,v)\big)  
    =
    \dve \Big(
        \delta_1^{\deg u}\cdot (^\sigma\! u)\Big(\frac{x}{\delta_1}\Big), 
        \delta_3(^\sigma\! v)\Big(\frac{x}{\delta_1}\Big) 
            + \delta_4\big(^\sigma\! (h\bmod u)\big)\Big(\frac{x}{\delta_1}\Big)
    \Big)
    \,.
\]

\paragraph{The case \(\deg u \le g \)}
Let us consider again the general case when \(\deg u \leq g\). 
Suppose \(\dve(u,v)\) is a divisor of maximal prime order \(r\)
(where \(r\) is a large prime factor of \(\#\Jac_\Hc(\F_q)\)
and \(r^2 \nmid \#\Jac_\Hc(\F_q)\)).
Then \(\Psi^\ast\) acts on \(\langle \dve(u,v) \rangle\) 
as multiplication by an eigenvalue \(0 \le \lambda < r\):
that is, \(\Psi^\ast(\dve(u,v)) = [\lambda] \dve(u,v)\).
We can compute \(\dve(u',v') \coloneqq [\lambda] \dve(u,v)\) using
Cantor's algorithm,
and then \(\dve(u^\ast,v^\ast)\coloneqq \Psi^\ast(\dve(u,v))\) must be equal to \(\dve(u',v')\). 

Write \(u'=\sum_i u'_ix^i\) and \(v'=\sum_i v'_ix^i\). 
Then \(u^\ast = u'\) and \(v^\ast = v'\) imply 
that \(\tilde{\delta_1} =\frac{1}{\delta_1},\delta_3,\delta_4 \in\F_q\) must belong to the varieties 
\[
    V_{1} / \F_q \colon
    \left\{\begin{array}{rl} 
    {\big(u_0\big)}^{2^\ell}  & = \big(\tilde{\delta}_1\big)^{\deg u} \cdot u'_0\\
    {\big(u_1\big)}^{2^\ell} \cdot \tilde{\delta}_1 & = \big(\tilde{\delta}_1\big)^{\deg u} \cdot u'_1\\
    {\big(u_2\big)}^{2^\ell} \cdot {\big(\tilde{\delta}_1\big)}^2 & = \big(\tilde{\delta}_1\big)^{\deg u} \cdot u'_2\\
    & \;\vdots  \\
    {\big(u_{\deg u - 1}\big)}^{2^\ell} \cdot{\big(\tilde{\delta}_1\big)}^{\deg u - 1} & = \big(\tilde{\delta}_1\big)^{\deg u} \cdot u'_{\deg u - 1}
    \end{array}\right.
\]
and 
\[
    \begin{split}
    V_{3,4} / \F_q \colon
    \left\{\begin{array}{rl} 
    \delta_3\cdot{\big(v_0\big)}^{2^\ell} + 
    \delta_4\cdot{\big(h'_0\big)}^{2^\ell} & = v'_0\\
    \delta_3\cdot{\big(v_1\big)}^{2^\ell}\cdot\tilde{\delta}_1 + 
    \delta_4\cdot{\big(h'_1\big)}^{2^\ell}\cdot\tilde{\delta}_1 & = v'_1\\
    \delta_3\cdot{\big(v_2\big)}^{2^\ell}\cdot{\big(\tilde{\delta}_1\big)}^2 + 
    \delta_4\cdot{\big(h'_2\big)}^{2^\ell}\cdot{\big(\tilde{\delta}_1\big)}^2 & = v'_2\\
    & \;\vdots \\
    \delta_3\cdot{\big(v_{\deg v}\big)}^{2^\ell}\cdot{\big(\tilde{\delta}_1\big)}^{\deg v} + 
    \delta_4\cdot{\big(h'_{\deg v}\big)}^{2^\ell}\cdot{\big(\tilde{\delta}_1\big)}^{\deg v} & = v'_{\deg v}
    \end{array}\right.
	\end{split}
\]

Observe that \(V_1\) only depends on the parameter \(\tilde{\delta}_1\), and 
it is determined by \((\deg u)\) polynomial equations of degree at most \(\deg u\). In particular, the \((\deg u)\)-th 
equation of \(V_1\) implies \(\tilde{\delta}_1 =\frac{{(u_{\deg u - 1})}^{2^\ell}}{u'_{\deg_u - 1}}\), if \(u_{\deg u - 1} \neq 0\). 
Otherwise, the \(i\)-th 
and \(j\)-th equations of \(V_{1}\) with \(j<i\) imply \(\tilde{\delta_1}^{i-j} = \frac{{(u_j)}^{2^\ell}\cdot u_i'}{{(u_i)}^{2^\ell}\cdot u_j'}\) 
when \(u_i\cdot u_j \neq 0\). 

The variety \(V_{3,4}/\F_q\) only depends on the parameters \(\delta_3\)  and \(\delta_4\), and 
it is determined by \((\deg v + 1)\) linear equations;
in fact, \(V_{3,4}(\F_q)\) consists of a unique point 
\((\delta_3,\delta_4)\in\F_q\times\F_q\). 
Combining the \(i\)-th and \(j\)-th equations of \(V_{3,4}\) 
yields
\begin{eqnarray}
	\delta_3 &=& \frac{v_i'\cdot{(\delta_1)}^i + \delta_4\cdot {(h_i')}^{2^\ell}}{{(v_i)}^{2^\ell}}
    \text{ and}
	\label{eq:d3}\\
	\delta_4	&=& \frac{ v_i'\cdot {(\delta_1)}^i\cdot {(v_j)}^{2^\ell} + v_j'\cdot {(\delta_1)}^j\cdot {(v_i)}^{2^\ell} }{
	{(h_i')}^{2^\ell}\cdot {(v_j)}^{2^\ell} + {(h_j')}^{2^\ell}\cdot {(v_i)}^{2^\ell}}
 \label{eq:d4}
\end{eqnarray}
where the denominators of Equations~\eqref{eq:d3} and~\eqref{eq:d4} are
different from zero. 

\section{Speeding-up the Index-Calculus algorithm in $\Jac_{\Hc}(\F_q)$}\label{sec:speedingup}

We now focus on the application of $\Psi^\ast$
to index calculus in \(\Jac_\Hc(\F_q)\).

\subsection{The speed-up in theory}
Recall from~\S\ref{sec:index-calculus}
that there are two main steps in index calculus.
First, the \textbf{relation generation} step:
having fixed a smoothness bound \(s\),
we generate $F(s) + \epsilon$ relations of the form
$\alpha_i \Dv + \beta_i\Dv'~=~\sum_{j=1}^{F(s)} m_{i,j}\Dv_j$,
where \(\Dv' = \lambda\Dv\) is the target DLP,
the \(\Dv_j\) are irreducible divisors of degree \(\leq s\),
where \(F(s)\) is the number of irreducible divisors \(\dve(u,v)\) with
\(\deg u \leq s\).
For the \textbf{linear algebra} step,
we construct the matrices
\(\alpha = (\alpha_i)^\mathsf{T}\), \(\beta = (\beta_i)^\mathsf{T}\) and
\(M = (m_{i,j})\) with coefficients in \(\Z/r\Z\);
the discrete logarithm can be recovered from a kernel vector \(\gamma\) of \(M\).

From~\S\ref{subsec:cost:index}, relation generation requires
$T(s) = (F(s) + \epsilon)E(s)$ random-walk steps, where $E(s)$ denotes the expected number of 
steps before an $s$-smooth divisor is found.
The kernel computation in the linear algebra step
requires $L(s) \approx d\cdot(F(s) + \epsilon)^2$ field operations.

Since the eigenvalue $\lambda$ of $\Psi^\ast$ satisfies $\lambda \ne \mp 1$ and 
$\lambda^n \pm 1 \equiv 0 \mod r$,
the divisors $\Dv, [\lambda]\Dv,\ldots,[\lambda^{n-2}]\Dv$ 
and $[\lambda^{n-1}]\Dv$ 
must be linearly dependent. 
Hence, whenever an \(s\)-smooth divisor is found, the endomorphism $\Psi^\ast$ allows us to 
obtain up to $n-1$ more $s$-smooth divisors
at essentially no cost.
However, the kernel vector $\gamma$  could 
produce the undesirable situation $\gamma \cdot \alpha \equiv 0$ and
$\gamma\cdot\beta \equiv 0$.
In order to prevent this, it seems more prudent to use only
\(n - 1\) related divisors, namely, $\Dv, [\lambda]\Dv,\ldots,[\lambda^{n-3}]\Dv$, and $[\lambda^{n-2}]\Dv$.
In other words, using $\Psi^\ast$ reduces the cost of relation generation
from $T(s)$ to just $\frac{T(s)}{n-1}$.

We can do even better by exploiting \(\Psi^\ast\) to reduce the factor
base size from \(F(s)\) to \(\frac{F(s)}{n}\). 
Mathematically speaking,
we work with the quotient of $\Jac_{\Hc}(\F_q)$ by the action of $\Psi^\ast$, 
taking a factor base of irreducible $\dve(u,v) \in \Jac_H(\F_q)$ with $\deg u \leq s$ consisting of
\(\Psi^\ast\)-orbit representatives.
(For example, the orbit of \(\dve(u,v)\) might be represented
by $\max\big( \{{(\Psi^\ast)}^i ( \dve(u,v) ) \colon 0 \leq i < n-1\}\big)$
with respect to the lexicographic ordering.) 
In other words,
$s$-smooth divisors factorize in our factor base as $\sum_i [{(\lambda^{j_i})}^{-1}]D_i$, where
$[\lambda^{j_i}]D_i = {(\Psi^\ast)}^{j_i}(D_i)$ is in the factor base for some $0 \le j_i < n$.
Using $\Psi^\ast$-orbits lets us reduce the costs $T(s)$ and $L(s)$ to
$\left(\frac{F(s)}{n} + \epsilon\right)E(s) \approx \frac{T(s)}{n}$ and
$d\cdot\left(\frac{F(s)}{n} + \epsilon\right)^2 \approx \frac{L(s)}{n^2}$, respectively.

While the factor-$n^2$ speed-up in the linear algebra phase is more
impressive than the factor-$n$ speed-up in relation generation,
linear algebra is not the bottleneck in the entire DLP computation.
The overall speed-up mostly corresponds to the speed-up in
the relation generation phase,
though the reduction in factor-base size
(and the resulting reduction in dimension of the linear algebra problem)
is still a very welcome improvement in practice.

\subsection{Problem instance: Solving discrete logarithms on \(\Ec / \F_{2^{5\times 31}}\)}\label{sec:toy}
In order to put the analysis above into practice,
we solved the DLP on an weak GLS binary curve
over \(\F_{q^\ell}\) where $q = 2^n$ with $n =5$ and $\ell = 31$.

Let
$\F_{q} = \F_{2}[u] / \langle u^5 + u^2 + 1 \rangle$ 
and $\F_{{q}^{\ell}} = \F_{q}[v] / \langle v^{31} + v^3 + 1\rangle$. 
The curve
\[ \Ec / \F_{q^{\ell}} \colon y^2 + x\cdot y = 
x^3 + x^2 + \big(v^{18} + v^{17} + v^{12} + v^8 + v^5 + v^4 + 1\big)
\]
satisfies $\# \Ec \left(\F_{{q}^{\ell}}\right) = c \cdot r$
where 
\(r = \mathrm{35153273567655620601556620437925421}\) is a $115$-bit prime number 
and \(c = \mathrm{1299222562550}\).

To construct a discrete logarithm challenge, we randomly selected an order-$r$ point 
$\Pt = (X_{\Pt}, Y_{\Pt})$ using the \texttt{Random()} function of Magma, and we 
set $\Pt' = \left[ c \right]\left( \pi_x, \pi_y \right)$ where 
\(\pi_x = v^{355}/v^{133} + (v + u + 1)\) and \(\pi_y\) is one of the roots 
of \(y^2 + \pi_xy + \pi_x^3 + \pi_x^2 + \big(v^{18} + v^{17} + v^{12} + v^8 + v^5 + v^4 + 1\big)\). 
Our goal was to find \(1 \le \lambda \le r\)
such that $\Pt' = [\lambda]\Pt$.
Magma code to set up this DLP instance is given in~\ref{App:E}.

Using the function \texttt{WeilDescent()} of Magma, we reduced the problem into a hyperelliptic genus-32 curve 
\(\Hc / \F_q: y^2 + h(x)y = f(x)\)
where
\begin{align*} 
    h(x) & = u^7x^{32} + u^{12}x^{16} + u^{30}x^8 + u^{28}x^2 + u^7x
    \,,
    \\
    f(x) & = u^4x^{65} + u^{14}x^{64} + u^{14}x^{33} + u^{19}x^{17} + u^{16}x^8 \\
    & \qquad \qquad \qquad \qquad \qquad \qquad {} + u^{15}x^5 + u^{25}x^4 + u^4x^3 + u^{24}x.
\end{align*}
The points \(\Pt\) and \(\Pt'\) are mapped to divisors \(\Dv\) and \(\Dv'\), respectively. 
The translated DLP instance in the Jacobian $\Hc / \F_{q}$ 
is described by the Magma code in~\ref{App:J}.
The endomorphism 
\(\Psi^\ast\) of \(\Jac_\Hc(\F_{q})\)
is defined by
\begin{equation*}
    \Psi^\ast \colon \dve(u,v) 
    \longmapsto 
    \dve \Big( {\big(u^{21}\big)}^{\deg u}\cdot (^\sigma\! u)\Big(\frac{x}{u^{21}}\Big), 
    \big(u^{14}\big)\cdot(^\sigma\! v)\Big(\frac{x}{u^{21}}\Big)\Big).
\end{equation*}

In this setting, we implemented a parallel version of the Enge--Gaudry algorithm in Magma~\cite{Magma}.
We successfully accelerated the relation step of the Enge--Gaudry algorithm by using the endomorphism \(\Psi^\ast\) as 
discussed in~\S\ref{sec:new_endom}. 
The factor base was dynamically built as in \cite{CO15}, using the smoothness 
bound \(s = 4\). 
The \(i\)-th thread of our parallel implementation built its own local factor base as
\begin{align*} 
    \mathcal{F}_{i,s} 
    = 
    \Big\{ \max \big\{ {(\Psi^\ast)}^i ( \dve(u,v) ) \colon 0 \le i < n \big\} \colon 
    \dve(u,v) \text{ irreducible},
    \deg u \leq s \Big\}
    \,.
\end{align*}
We used 96 cores of 16 Intel Core i7 machines (3.20GHz, 3.40GHz, and 3.47GHz) and 
32 cores of two Intel Xeon E5 2.60GHz machines to find our 4-smooth divisors.
The linear algebra step was solved with one core of an Intel Xeon E5 2.60GHz machine by using 
the function \texttt{ModularSolution()} that Magma has implemented. 
We solved the DLP in \(\Jac_\Hc(\F_{2^5})\) in 1034.596 CPU days,
finding
\[
    \gamma 
    = 
    31651293342165466420895111254857443
    \,.
\]
The Magma implementation of the procedures described here is available at
\url{https://github.com/JJChiDguez/combining_GLS_with_GHS.git}.

\subsection{Comparison with related work}
Velichka, Jacobson, and Stein~\cite{VJS14}
report the solution of a discrete logarithm problem for the same
elliptic curve \(\Ec / \F_{2^{5\times 31}}\)
using hyperelliptic index calculus without the endomorphism technique.
Table~\ref{tb:exp:comp} compares our results with theirs.

\begin{table}[t]
    \centering
    \begin{tabular}{|l|c||c|c|c|}
        \hline
        \multirow{2}{*}{}  & \multirow{2}{*}{This work} & \multicolumn{3}{c|}{Velichka \textit{et al.} \cite{VJS14}} \\
        \cline{3-5}
        & & JMS EG & Opt.~EG & Vollmer
        \\
        \hline
        \hline
        Relation generation & \textbf{1034.572} & \emph{8492.67} & \emph{6338.01} & 1720.818 \\
        \hline
        Linear algebra step & \textbf{0.024} & \emph{2.470} & \emph{2.800} & 14.244 \\
        \hline
        \hline
        Total & \textbf{1034.597} & \emph{8495.650} & \emph{6340.810} & 1735.063 \\
        \hline
        \hline
        \multicolumn{2}{|l||}{\textit{Speedup}} &
        \emph{\textbf{8.212}} & \textbf{\emph{6.129}} & \textbf{1.677} \\
        \hline
    \end{tabular}
    \caption{CPU days to
    solve the DLP on tbe hyperelliptic genus-32 curve $\Hc / \F_{2^5}$
    of~\S\ref{sec:toy},
    using index calculus with smoothness bound \(4\).
    Values in parentheses are estimates.
    ``JMS EG'' and ``Opt. EG''
    are estimates from~\cite{VJS14} for
    the Enge--Gaudry algorithm with the strategy and optimal parameters from~\cite{JMS01},
    and an optimized large-prime variant,
    respectively.
    ``Vollmer'' lists experimental timings from~\cite{VJS14} 
    using a sieve-based version of Vollmer's algorithm.
    }
    \label{tb:exp:comp}
\end{table}

The factor base in~\cite{VJS14} had \(136,533\) divisors.
Using the endomorphism described here,
we reduced this to \(27271 \approx \frac{136533}{5}\) divisors, 
in line with our theoretical analysis.

The discrete logarithm in~\cite{VJS14}
was computed using a sieve-based version of Vollmer's algorithm
implemented with the GNU Multi-Precision C library version 4.2.2, 
Automatically Tuned Linear Algebra Software (ATLAS) version 3.7.31 ([2]), 
and linbox version 1.1.3 compiled with GCC version 3.4.4 for the linear
algebra.  Their experiments were run on 
152 dual Intel P4 Xeon machines (2.4GHz and 2.8GHz) 
with 512 kb cache and 2 GB of RAM.
We also find estimated timings
in~\cite{VJS14} for hypothetical computations using the Enge--Gaudry
algorithm using parameters derived from~\cite{JMS01},
and for a large-prime variation.

Remarkably, we managed to produce a faster discrete logarithm attack than the one reported in~\cite{VJS14},
despite using a non-optimized implementation based on Magma. Due to the more advanced micro-architecture used 
in our experiments, the speedup achieved by our approach was higher than expected.\footnote{It is worth mentioning 
that Magma's implementation of Lanczos algorithm takes advantage of both, a more advanced micro-architecture 
instruction set and a concurrent multi-core computation.}

The endomorphism  \(\Psi^\ast \colon \Jac_\Hc(\F_q) \to \Jac_\Hc(\F_q)\) can also be 
used in the sieve-based  version of Vollmer's algorithm~\cite{VJS14}. Extrapolating the timing costs given in 
Table~\ref{tb:exp:comp}, we would expect 344.164 and 0.569 CPU-days 
for the relation generation and linear algebra steps, respectively.

\section{Conclusions}\label{sec:concl}
We have shown that the GLS endomorphism on \(\Ec / \F_{2^{n\cdot\ell}}\) induces an 
efficient endomorphism \(\Psi^\ast \colon \Jac_\Hc(\F_q) \to \Jac_\Hc(\F_q)\) on the 
Jacobian of the image of GHS Weil descent applied to \(\Ec / \F_{2^{n\cdot\ell}}\). 
This endomorphism permits a factor-$n$ speedup over standard 
index-calculus procedures for solving the DLP on \(\Jac_\Hc(\F_q)\). 
Our analysis is backed up by the explicit computation of a DLP in a prime-order 
subgroup of a GLS elliptic curve over the field $\F_{2^{5\cdot 31}}$.
A Magma implementation of a standard index-calculus procedure boosted
with the GLS endomorphism found this discrete logarithm in about $1,035$ CPU-days.

While binary GLS curves offer a tempting speedup for scalar
multiplication,
our results show that this is tempered by a substantial speedup in DLP
computations.
This must be taken into account if binary GLS curves are considered for
use in cryptographic applications.

\section*{Acknowledgement}
The authors would like to acknowledge the anonymous referees whose comments and suggestions greatly helped us to
improve the  manuscript.
We thank ``Consejo Nacional de Ciencia y Tecnolog\'ia'' (CONACyT) for the scholarship they provided
to the first author during the period that he was a Ph.D. candidate at the Computer Science Department
of Cinvestav-IPN.

\bibliography{mybibfile}

\appendix
\section{Magma codes}
\subsection{Elliptic curve instances: EC\_instance.mag}
\label{App:E}
{\footnotesize
\begin{verbatim}
n := 5; l := 31; q := 2^n; N := 2^l;
F_2 := GF(2); P_2<t> := PolynomialRing(F_2);

F_q<u> := ext<F_2| t^5 + t^2 + 1>;
F_qn<v>:= ext<F_q| t^31 + t^3 + 1>;

a_qn := F_qn!1; b_qn := v^18 + v^17 + v^12 + v^8 + v^5 + v^4 + 1;
E_qn := EllipticCurve([F_qn| 1, a_qn, 0, 0, b_qn]);
c := 1299222562550; r := 35153273567655620601556620437925421;

Pt_x := F_qn![ u^10, u^30, u^24, u^17, u^26, u^23, u^22, u^8, 
u^4, u^25, u^24, u^19, 0, u^30, u^2, u^8, u^24, u^16, u^21, 
u^19, u^3, u^2, u^21, u^7, u^11, u^4, u^23, u^13, u^3, u^23, u^23 ];
Pt_y := F_qn![ u^25, u^29, u^16, u^20, 0, 1, u^10, u^6, u^13, 
u^30, u^8, u^30, u^9, u^9, 0, u^9, u^8, u^28, u^21, u^23, u^23, 
u^16, u^27, u^22, u^8, u^4, u^8, u^12, u^17, u^7, u^9 ];
Pt := E_qn![Pt_x, Pt_y];

Pt_prime_x := v^355/v^133 + (v+u+1);
Pt_prime_y := F_qn![ u^15, u^12, u^12, 1, u^15, u^22, u^16, 0, 
u^17, u^3, u^19, u^10, u^9, u^25, u^18, u^23, u^13, u^9, u^12, 
u^22, u^30, u^17, u^15, u^22, u^2, u^22, u^21, u^16, u^13, u^7, u^20 ];
Pt_prime := c*E_qn![Pt_prime_x, Pt_prime_y];
\end{verbatim}
}
\subsection{Hyperelliptic curve instances: HEC\_instance.mag}
\label{App:J}
{\footnotesize
\begin{verbatim}
P_q<w> := PolynomialRing(F_q);
h_q := u^7*w^32 + u^12*w^16 + u^30*w^8 + u^28*w^2 + u^7*w;
f_q := u^4*w^65 + u^14*w^64 + u^14*w^33 + u^19*w^17 + u^16*w^8 
+ u^15*w^5 + u^25*w^4 + u^4*w^3 + u^24*w;

H_q := HyperellipticCurve(f_q, h_q);
J_q := Jacobian(H_q);

D_x := P_q![ u^9, u^18, u^28, u^3, u^29, u^21, u^17, u^19, u^26, 
u^16, u^8, u^25, u^11, u^8, u^5, u^18, 0, u^2, u^21, u^3, u^28, 
u^19, u^22, u^14, u^24, u^6, u^28, u^19, u^16, u^21, u^20, u^18, 1 ];
D_y := P_q![ u^4, u^24, 0, u^2, u^20, u^18, u^30, u, u^6, u^6, 
u^27, u^29, u^14, u^29, u^17, u^10, u^12, u^23, u^11, u^3, u^12, 
u^11, u^9, u^14, u^30, u^25, u^6, 0, u^5, u^2, u^29, u^25 ];
D := J_q![D_x, D_y];

D_prime_x := P_q![ u^19, u^8, u^23, u^7, u^26, 0, u^2, u^4, u^21, 
u^12, u^17, u^20, u^22, u^2, u^5, u^17, u, u^27, u^28, u^16, u^6, 
u^18, u^5, u^27, u^19, u^15, u^11, u^14, u^8, u^6, u^26, u^11, 1 ];
D_prime_y := P_q![ u^2, u^24, u^21, u^13, u^10, u^17, 1, u^15, 
u^29, u^3, u^16, u^4, u, u^17, u^13, u^22, u^26, u^18, u^8, u^16, 
u^21, u^26, u, u^16, u^16, u^3, u^5, u^24, u^26, u^26, u^14, u^14];
D_prime := J_q![D_prime_x, D_prime_y];
\end{verbatim}
}
\subsection{Testing the solution: checking\_dlog.mag}
\label{App:dLog}
{\footnotesize
\begin{verbatim}
load "EC_instance.mag";
load "HEC_instance.mag";

dLog := 0x618877C96DE350E8C7980393356E3;
(Pt * dLog) eq Pt_prime; (D * dLog) eq D_prime;
\end{verbatim}
}
\end{document}